\documentstyle[amssymb,pre,aps,preprint]{revtex}

\begin{document}
\title{Two point Correlations and Critical Line of the Driven Ising Lattice Gas in
a High Temperature Expansion}
\author{B. Schmittmann and R. K. P. Zia}
\address{Center for Stochastic Processes in Science and Engineering \\
and \\
Department of Physics, Virginia Polytechnic Institute and State \\
University \\
Blacksburg, Virginia 24061-0435, USA}
\date{March 3, 1998}
\maketitle

\begin{abstract}
Based on a high temperature expansion, we compute the two-point correlation
function and the critical line of an Ising lattice gas driven into a
non-equilibrium steady state by a uniform bias $E$. The lowest nontrivial
order already reproduces the key features, i.e., the discontinuity
singularity of the structure factor and the (qualitative) $E$-dependence of
the critical line. Our approach is easily generalized to other
non-equilibrium lattice models and provides a simple $analytic$ tool for the
study of the high temperature phase and its boundaries.
\end{abstract}

\section{Introduction}

The study of non-equilibrium steady states (NESS) has attracted vivid
interest over the past decade. On the one hand, such studies are
application-driven, since NESS determine the physics of a wide range of
important problems, including, e.g., granular and traffic flow, surface
growth, electromigration, and transport phenomena in biological systems. On
the other hand, there is a fundamental interest in creating a theoretical
framework for NESS on a par with Gibbs ensemble theory for equilibrium
systems. Driven diffusive lattice gases, first introduced by Katz, et.al. 
\cite{KLS} and recently reviewed in \cite{SZ}, provide simple testing
grounds for the properties of a particular class of NESS. Based mostly on
Ising lattice gases with Kawasaki (spin exchange) dynamics, these models are
represented by microscopic Master equations which violate the usual detailed
balance condition by maintaining a net probability current between
configurations. A particularly simple member of this class is the uniformly
driven lattice gas, in which spin exchanges along a specific lattice axis
are biased by a uniform force, remaining energetically controlled only in
the transverse subspace. Identifying the particles with (shielded) charges
and the force with an electric field, such models provide a starting point
for the description of fast ionic conductors\cite{FIC} or charged droplets
in a microemulsion \cite{ME}.

One of the more remarkable features of this model, and many of its
relatives, is the presence of long-range correlations at all temperatures 
\cite{ZWLV,ZHSL}. Though this behavior can be understood, within the context
of a phenomenological (mean) field theory \cite{ZHSL}, in terms of the
violation of the traditional fluctuation-dissipation theorem \cite{FDT}, it
is clearly important to have an exact microscopic version. Using a venerable
tool, the high temperature series expansion, Zhang, et.al. \cite{ZWLV} (ZWLV
in the following) investigated $G,$ the two-point correlation function for a
system with infinite $E$. From the Master equation and the associated
hierarchy, they argued that three-point correlations are negligible and
arrived at a closed set of equations for $G$ alone. To zeroth order in $%
\beta $, both the equation and the solution are trivial. To first order, the
short-distance behavior of the solution was obtained numerically, by
truncating the equations at distances larger than a cutoff value. The
results exhibit significant anisotropy and agree quite well with Monte Carlo
data at high temperatures, i.e., $T\gtrsim 6$, in units of $T_{c}(0)$, the
critical temperature of the $E=0$ system. Analytically, the equations were
approximated by a Poisson problem with quadrupole symmetry. Its solution
captures the behavior of $G$ for large distances, displaying the $r^{-2}$
power law tail, with the angle-dependent amplitude of the quadrupole
potential, in agreement with simulations. We should emphasize that the high
temperature series has a strong foundation: for $\beta J\equiv 0$, the
steady state distribution $P^{*}$ $\propto 1$ is exactly known for all $E$ 
\cite{Spitzer}, so that we are expanding about a well-defined zeroth order
state. In this paper, we generalize the analysis of ZWLV to include finite
driving fields and solve the resulting set of equations for $G$ {\em exactly}%
, by computing its Fourier transform, the structure factor $S$. The latter
displays the characteristic discontinuity singularity \cite{ZHSL,KLS} at the
origin, which translates into the $r^{-d}$ decay in real ($d$-dimensional)
space. Thus, recourse to numerical methods is not necessary.

Another key feature of driven lattice gases is the existence of a line of
continuous transitions, falling outside the Ising universality class \cite
{crit}. In particular, it is remarkable that the critical temperature, $%
T_{c}(E)$,{\em \ increases} with $E$, even though the drive reduces the
effective nearest-neighbor coupling. Apparently, the strong long-range part
of the correlations dominates the driven system and permits the onset of
order at a higher $T_{c}$ \cite{HZS}. To support this argument, it is
desirable to have a simple analytic method to estimate {\em both}
correlations {\em and} the critical temperature. The high temperature
expansion, and specifically our explicit form for the structure factor,
serve that purpose. Matching the series expansion of $S$ to the expected
critical singularity, we obtain estimates for the critical line, and hence
the phase diagram to this (lowest) order in $\beta $.

Clearly, quantitative accuracy cannot be expected from a first-order
calculation. Nevertheless, it presents one of the {\em simplest, currently
available, tools} for the qualitative prediction of nonuniversal quantities
in driven lattice gases. Real-space renormalization group techniques for
conserved non-equilibrium spin systems are sorely lacking, and dynamic
mean-field theories\cite{DMF}, while quantitatively more satisfying, are
much more labor-intensive. Our method is easily extended to general rate
functions, anisotropic interactions and higher dimensions. These results
will be published elsewhere \cite{long}.

The paper is organized as follows: we first summarize the model definition
and the method of ZWLV, resulting in a closed set of equations for the
two-point function $G$, correct to first order in $\beta J$ but for
arbitrary drive strength $E>12J$. These equations are then solved exactly in
Fourier space. Key consequences, such as the discontinuity singularity of
the structure factor and the critical line $T_{c}(E)$, are discussed. We
conclude with a brief comment on results in higher dimensions.

\section{The Model and the Equations for its Two-point Function}

We first summarize the microscopics of the model. On each site $\vec{r}$ of
an (infinite) square lattice in spatial dimension $d=2$, coupled to a heat
bath at inverse temperature $\beta $, we define an Ising spin variable $S_{%
\vec{r}}$ which distinguishes occupied ($S=+1$) from empty ($S=-1$) sites.
Interacting via the usual Ising Hamiltonian, ${\cal H=}-J\sum S_{\vec{r}}S_{%
\vec{r}^{\prime }}$, nearest neighbors can exchange positions, subject to
the local energetics and a uniform force $E$ (``electric field'') which
biases exchanges along a specific lattice axis (aligned with the $x$%
-coordinate). To be specific, we choose Metropolis \cite{Metropolis} rates, $%
\min \{1,\exp -\beta [\Delta {\cal H}-\sigma E]\}$, where $\sigma =0,+1,-1$
for jumps transverse, along and against the field. Thus, ``infinite'' $E$
implies that jumps along the field always take place, provided they are
allowed by the excluded volume constraint, while jumps against $E$ are
completely suppressed. These rates specify the master equation of the model,
for the time-dependent configurational probability $P$. A hierarchy of
equations, connecting different $N$-point functions, follows as usual. We
will be interested in steady-state averages only, so all time derivatives
will be set to zero. Following ZWLV, we expand the rate functions in powers
of $\beta J$; as an extension of their work, which focused on $E=\infty $
only, we include the effect of large, but {\em finite} $\beta E$. Thus, in a
sense, the case here corresponds to $\beta E\gg \beta J$. In practice, it is
sufficient to choose $E>12J$, so that jumps along the drive occur with unit
rate, while those against $E$ are suppressed by a factor of $\exp (-\beta E)$%
. In a strict high temperature expansion, organized in powers of $\beta $, $%
\beta E$ would also appear as a small parameter, so that the range $E<12J$
can also be explored \cite{long}.

Returning to our case, a set of equations for the two-point function, $G(%
\vec{r},\vec{r}^{\prime })\equiv <S_{\vec{r}}S_{\vec{r}^{\prime }}>$, is
easily derived from the master equation with the expanded rates. At first
order in $\beta J$, one finds that $G$ couples only to three-point
functions. Even though, in contrast to the equilibrium Ising model, the
latter do not vanish here \cite{3pf}, they are numerically quite small and
are neglected. A closed set of equations for $G$ emerges. By translational
invariance, $G$ depends only on $(x,y)$, the separation between the two
points $\vec{r}$ and $\vec{r}^{\prime }$, and it is even in both variables.

The resulting equations depend on two parameters, 
\[
K\equiv \beta J\quad \text{and}\quad \epsilon \equiv \exp (-\beta E)\quad . 
\]
Up to and including only first order terms in $K$, we obtain a set of linear
equations for $G$, 
\begin{eqnarray}
0 &=&2[G(1,1)+G(1,-1)-2G(1,0)]+(1+\epsilon )[G(2,0)-G(1,0)]+2K(2+\epsilon ) 
\nonumber \\
0 &=&2[G(0,2)-G(0,1)]+(1+\epsilon )[G(1,1)+G(-1,1)-2G(0,1)]+2K(1+2\epsilon )
\nonumber \\
0 &=&2[G(1,2)+G(1,0)-2G(1,1)]+(1+\epsilon
)[G(2,1)+G(0,1)-2G(1,1)]-2K(1+\epsilon )  \label{Gnr0} \\
0 &=&2[G(2,1)+G(2,-1)-2G(2,0)]+(1+\epsilon
)[G(3,0)+G(1,0)-2G(2,0)]-2\epsilon K  \nonumber \\
0 &=&2[G(0,3)+G(0,1)-2G(0,2)]+(1+\epsilon )[G(1,2)+G(-1,2)-2G(0,2)]-2K 
\nonumber
\end{eqnarray}
and, for all other non-zero $x,y$: 
\begin{eqnarray}
0 &=&2[G(x,y+1)+G(x,y-1)-2G(x,y)]  \nonumber \\
&&+(1+\epsilon )[G(x+1,y)+G(x-1,y)-2G(x,y)]  \label{Gfar}
\end{eqnarray}
Here, we have written the equations in such a form that the left hand side
would simply be $\partial _{t}G(x,y)$. The ZWLV equations can be retrieved
by setting $\epsilon =0.$ The solution to these equations is most easily
found in Fourier space. In other words, we consider the structure factor: 
\[
S(k,p)=\sum_{x,y}^{\infty }G(x,y)\exp [-i(kx+py)] 
\]
which is real, since $G$ is even. Some details on Fourier transforms are
given in Appendix A.

In the absence of $J$, the solution is trivial: 
\begin{eqnarray*}
G(0,0) &=&1 \\
G(x,y) &=&0\ \ \text{ }x,y\neq 0\text{,}
\end{eqnarray*}
so that 
\[
S(k,p)=1\equiv \bar{S} 
\]
reflecting, of course, that $P^{*}\propto 1$ in this case\cite{Spitzer}. So,
the information about interactions is carried by $\tilde{S}$, defined by: 
\[
S=\bar{S}+\tilde{S}\quad . 
\]
The correlations appearing in (\ref{Gnr0}), being proportional to $K$, are
really the transforms of $\tilde{S}$: 
\begin{equation}
G(x,y)=\int \ \tilde{S}(k,p)\ \exp i(kx+py)\quad ,\text{ }(x,y)\neq (0,0)
\label{G-ft}
\end{equation}
For convenience, we~have introduced the notation 
\[
\int \equiv \frac{1}{(2\pi )^{2}}\int_{-\pi }^{+\pi }dk\int_{-\pi }^{+\pi
}dp 
\]
and will denote the anisotropic lattice Laplacian by 
\[
\Delta (k,p)\equiv 2(1+\epsilon )(1-\cos k)+4(1-\cos p)\quad . 
\]

Inserting (\ref{G-ft}) into (\ref{Gnr0}, \ref{Gfar}) and using the fact that 
$\tilde{S}$ is real, we obtain 
\begin{eqnarray}
\int \tilde{S}\Delta \exp (ik)+(1+\epsilon )\int \tilde{S}(1-\cos k)
&=&+2(2+\epsilon )K  \nonumber \\
\int \tilde{S}\Delta \exp (ip)+2\int \tilde{S}(1-\cos p) &=&+2(1+2\epsilon )K
\nonumber \\
\int \tilde{S}\Delta \exp [i(k+p)] &=&-2(1+\epsilon )K  \label{Snr0} \\
\int \tilde{S}\Delta \exp (2ik) &=&-2\epsilon K  \nonumber \\
\int \tilde{S}\Delta \exp (2ip) &=&-2K  \nonumber
\end{eqnarray}
with similar equations for negative values of $x$ or $y$ and 
\begin{equation}
\int \tilde{S}\Delta \exp [i(kx+py)]=0\text{ \ \ for }|x|+|y|>2\text{ .}
\label{Sfar}
\end{equation}

Next, we seek to invoke the completeness relation (see the Appendix) in
order to project out an equation for $\tilde{S}$. While the two additional
terms on the left hand side of (\ref{Snr0}) might appear to spoil this
approach, we can just treat them, for the time being, as unknown $\epsilon $%
-dependent coefficients, 
\begin{eqnarray}
I_{1} &\equiv &\int \tilde{S}(1-\cos k)  \nonumber \\
I_{2} &\equiv &\int \tilde{S}(1-\cos p)  \label{I's}
\end{eqnarray}
and move them to the right hand side. Finally, for completeness, we need an
additional equation for $x=y=0$, namely: 
\begin{equation}
\int \tilde{S}\Delta =\int \tilde{S}\ [2(1+\epsilon )(1-\cos k)+4(1-\cos
p)]=2(1+\epsilon )I_{1}+4I_{2}\quad .  \label{0,0}
\end{equation}

Now that we have equations for {\em all} integer values of $x,y$, we can use 
$\sum_{x,y}\exp [i(kx+py)]=(2\pi )^{2}\delta (k)\delta (p)$. The result is 
\begin{equation}
\tilde{S}(k,p)=L(k,p)/\Delta (k,p)  \label{Stilde}
\end{equation}
where $L$ is the sum of terms on the right hand side, i.e., 
\begin{eqnarray}
L(k,p) &=&2I_{1}(1+\epsilon )(1-\cos k)+4I_{2}(1-\cos p)  \nonumber \\
&&+4K\left[ \epsilon (1-\cos k)+(1-\cos p)\right] \left[ 1+2\cos k+2\cos
p\right] \quad .  \label{LKP}
\end{eqnarray}

Of course, (\ref{Stilde}) is still an implicit equation for the structure
factor, due to the appearance of $I_{1}$ and $I_{2}$ on the right. To find
the explicit solution, we must determine the $I$'s. Since we have two
unknowns, we need two linearly independent equations. The first of these
follows from (\ref{I's}): Inserting our result for $L(k,p)$, Eqn. (\ref{LKP}%
) into the first equation of (\ref{I's}), we obtain 
\[
0=-I_{1}+\int \frac{L(k,p)}{\Delta (k,p)}(1-\cos k)=M_{1,j}I_{j}+KN_{1} 
\]
where 
\begin{eqnarray*}
M_{1,1} &=&-1+2(1+\epsilon )\int (1-\cos k)^{2}/\Delta (k,p) \\
M_{1,2} &=&4\int (1-\cos k)(1-\cos p)/\Delta (k,p)
\end{eqnarray*}
and 
\[
N_{1}=\int \left( 1-\cos k\right) \left[ 4(1-\cos p)+4\epsilon (1-\cos
k)\right] \left[ 2\cos k+2\cos p+1\right] /\Delta (k,p) 
\]

Due to a remarkable symmetry under $(k,p)$ exchange \cite{long}, the second
equation in (\ref{I's}) is {\em not} linearly independent from the first.
Instead, an additional equation is provided by the value of $G$ at the
origin, i.e., $1=G(0,0)=\int S(k,p)$ which leads to 
\[
0=\int \tilde{S}(k,p)=\int \frac{L(k,p)}{\Delta (k,p)}=M_{2,j}I_{j}+KN_{2} 
\]
with 
\begin{eqnarray*}
M_{2,1} &=&2(1+\epsilon )\int (1-\cos k)/\Delta (k,p) \\
M_{2,2} &=&4\int (1-\cos p)/\Delta (k,p)
\end{eqnarray*}
and 
\[
N_{2}=\int \left[ 4(1-\cos p)+4\epsilon (1-\cos k)\right] \left[ 2\cos
k+2\cos p+1\right] /\Delta (k,p)\text{ .} 
\]
Note that the singularity of $1/\Delta (k,p)$ at the origin is cancelled by
zeros in the numerators, so that all of these integrals are perfectly
finite. The explicit solution now follows as 
\begin{equation}
I_{m}=-KM_{mn}^{-1}N_{n}\text{ .}  \label{I's_exp}
\end{equation}
Together with (\ref{LKP}), this determines the full structure factor,
displaying the expected proportionality $\tilde{S}(k,p)\propto K$ .

For later reference, let us briefly consider the equilibrium analog to our
results so far. To first non-trivial order in $K$, the two-point correlation
of the Ising model is given by the well-known form \cite{HTS} 
\begin{eqnarray*}
G_{eq}(0,0) &=&1 \\
G_{eq}(1,0) &=&G_{eq}(0,1)=K \\
G_{eq}(x,y) &=&0\text{ \ for all other \ }x,y\text{ .}
\end{eqnarray*}
This results in a structure factor $S_{eq}(k,p)=1+2K(\cos k+\cos p)$ which
is of course isotropic.

\section{Discontinuity Singularity and Criticality}

In this section, we discuss two of the most interesting consequences of the
full solution, 
\begin{equation}
S=\bar{S}+\frac{L}{\Delta }+O(K^{2})\text{ .}  \label{Ans}
\end{equation}
First, we focus on the celebrated discontinuity of the structure factor near
the origin \cite{ZHSL,KLS}. This anomaly is a direct consequence of FDT
violation and is therefore expected to increase in magnitude with the
strength of the drive. For small $k,p$ we find 
\begin{eqnarray*}
\Delta  &=&(1+\epsilon )k^{2}+2p^{2}+O(k^{4},k^{2}p^{2},p^{4}) \\
L &=&[(1+\epsilon )I_{1}+10\epsilon
K]k^{2}+2[I_{2}+5K]p^{2}+O(k^{4},k^{2}p^{2},p^{4})
\end{eqnarray*}
Thus, the discontinuity can be measured by 
\[
\lim_{p\rightarrow 0}S(0,p)-\lim_{k\rightarrow 0}S(k,0)=I_{2}-I_{1}+5K\frac{%
1-\epsilon }{1+\epsilon }
\]
Of course, this is proportional to $K$. The dependence on $E$ is captured by
the parameter $\epsilon $, which also enters into the expressions for the
integrals $I_{1}$ and $I_{2}$ (see the Appendix for a discussion and some
characteristic values). Consistent with our expectation, we observe that the
discontinuity increases monotonically with $E$, from a limiting value of $0$
for $12K<\beta E\ll 1$ to $5.48$ for infinite drive.

Originally obtained from field-theoretic considerations, the form which best
displays this discontinuity is \cite{crit,SZ}, 
\begin{equation}
S(k,p)=\frac{n_{\Vert }k^{2}+n_{\bot }p^{2}}{\tau _{\Vert }k^{2}+\tau _{\bot
}p^{2}+O(k^{4},k^{2}p^{2},p^{4})\ }  \label{owl}
\end{equation}
near the origin. Here, $n_{\Vert }$ and $n_{\bot }$ measure the strength of
thermal noise in the parallel and transverse directions, respectively, while 
$\tau _{\Vert }$ and $\tau _{\bot }$ are the anisotropic diffusion
coefficients. Under equilibrium conditions, the FDT enforces the equality 
\[
\frac{n_{\Vert }}{\ n_{\bot }}=\frac{\tau _{\Vert }}{\tau _{\bot }\ } 
\]
so that the familiar Ornstein-Zernike form re-emerges. By contrast, we
conclude that 
\[
\frac{n_{\bot }}{\tau _{\bot }\ }-\frac{n_{\Vert }}{\tau _{\Vert }\ }%
=I_{2}-I_{1}+5K\frac{1-\epsilon }{1+\epsilon }\neq 0 
\]
in the driven case.

Next, we turn to an estimate for the critical temperature. If we had the
exact $S(k,p)$, we could identify $T_{c}$ by its divergence at some point.
For the usual system in equilibrium, we would look for the divergence of $%
S(0,0)$, which is the susceptibility ($\tau ^{-1}$). However, with conserved
dynamics at half-filling, this quantity is fixed at zero, so that it is
necessary to consider $\lim_{{\bf k\rightarrow 0}}S({\bf k})$. In our case,
as we just pointed out, this limit depends on the direction along which we
approach the origin. Thus, we appeal first to the phenomenology, i.e., only $%
\lim_{p\rightarrow 0}S(0,p)$ diverges as $T\rightarrow T_{c}$. In terms of (%
\ref{owl}), only $n_{\bot }/\tau _{\bot }$ $\rightarrow \infty $. In terms
of co-operative behavior, this corresponds to an instability against
ordering into strips {\em parallel} to the drive only.

Now, in a high temperature series, where only a finite number of terms can
be computed, every partial sum is finite. Instead, the radius of convergence
must be estimated. However, here we have only one non-trivial term! To make
any estimate, we turn to a search for the ${\em zero}$ of $S^{-1}$ instead.
For the equilibrium case, $S_{eq}^{-1}(k,p)=1-2K(\cos k+\cos p)+O(K^{2})$,
so that this procedure leads to $T_{c}=4J/k_{B}$. Of course, this is the
same result, had we expanded the exact equation for $\beta _{c}$ ($%
e^{-4\beta _{c}J}+2e^{-2\beta _{c}J}=1$) in powers of $\beta _{c}J$ and kept
only the first non-trivial term. Remarkably, this is also the mean field
critical temperature.

Turning to the driven case, we consider an expansion of $S^{-1}$ (given that 
$\bar{S}=1$): 
\begin{equation}
S^{-1}=1-\frac{L}{\Delta }+O(K^{2})\text{ ,}  \label{1/S}
\end{equation}
and seek the zero of $\lim_{p\rightarrow 0}S^{-1}(0,p)$. The result is 
\begin{equation}
T_{c}(E)=\left( J/k_{B}\right) \left[ 5+I_{2}/K\right] \text{ .}  \label{TcE}
\end{equation}
Recall that $I_{2}$ is proportional to $K$, so that their ratio is a $\beta $%
-independent number (Table 1).

Before discussing the implications of this result, let us identify (up to an
overall constant) the various parameters in Eqn. (\ref{owl}) using this
approach: 
\begin{eqnarray*}
&&\tau _{\parallel }=(1+\epsilon )(1-I_{1})-10K\epsilon \\
&&\tau _{\perp }=2(1-I_{2}-5K) \\
&&n_{\parallel }=1+\epsilon \\
&&n_{\perp }=2
\end{eqnarray*}
Referring to the discussion of $I_{1}$ and $I_{2}$ given in the Appendix, we
see that $\tau _{\parallel }>\tau _{\perp }$ for all $E>0$. This inequality
confirms our choice to identify the critical temperature by the vanishing of 
$\tau _{\perp }$.

Returning to (\ref{TcE}), let us discuss critical temperatures in units of $%
J/k_{B}$. First, note that $I_{2}$ is negative and monotonically decreasing
in $\epsilon =\exp (-\beta E)$ (cf. Appendix). Therefore, $T_{c}(E)$
increases with $E$, taking its maximum at infinite $E$ where $T_{c}(\infty
)=4.640$. For the smallest $\beta E$, defined through the inequality $%
12K<\beta E\ll 1$, $T_{c}(E)$ approaches its lowest value of $4.0$. The
agreement with the equilibrium result, $T_{c}(0)=4$, is an artifact of the 
{\em lowest} order of the expansion only \cite{long}. Clearly, it is
gratifying that even the lowest nontrivial order of the high-temperature
expansion generates a $T_{c}(E)$ which $increases$ with $E$, in {\em %
qualitative }agreement with simulation data. Finally, let us consider the 
{\em quantitative} implications of our results by focusing on the\ ratio $%
T_{c}(E)/T_{c}(0)$. For $E=\infty $, our approach yields the value $1.16$,
while MC simulations result in $1.40$ \cite{MC2}. Thus, the series
underestimates this ratio, which can be understood as follows. The high
temperature series is known to overestimate critical temperatures, by
underestimating fluctuations. However, it has been argued \cite{HZS} that
the external drive tends to suppress fluctuations, so that we may expect $%
T_{c}(\infty )$ to be {\em less} sensitive to the numerical errors
introduced by the high temperature series than its equilibrium counterpart.
In this sense, the series expansion should be ``better'' for a driven
system. Indeed, we compare the series result of $4.64$ to the simulation
value $3.18$, finding a discrepancy of $46\%$. In contrast, for the
equilibrium case, we have $4.00$ and $2.27$ respectively, showing a much
higher discrepancy of $76\%$.

Finally, we could hope for better agreement of series and the exact $T_{c}$
as we move into higher spatial dimensions, where fluctuations become less
important. The results are certainly encouraging in $d=3$. In the series
approach, we obtain $T_{c}(0)=6$ and $T_{c}(\infty )=6.34$, signalling an
increase of $6\%$ due to the drive. This is in remarkably good agreement
with the MC data, which show a $7\%$ increase \cite{MC3}. Thus, even a
low-order calculation can produce some quantitatively reliable results.

\section{Conclusions}

Within a high temperature series, we have derived the equations for the
two-point correlations of the uniformly driven lattice gas to lowest
nontrivial order in $\beta J$, but finite $\beta E$. The exact solution of
these equations provides a successful qualitative description of two central
features of our model, namely, the discontinuity singularity of the
structure factor at the origin, associated with power-law correlations in
the disordered phase, and the anisotropy in the parallel and transverse\/
diffusion coefficients which controls the onset of criticality.
Specifically, we observe that the magnitude of the structure factor
discontinuity increases with $E$, as a measure of how seriously the FDT is
violated in the driven system. We demonstrate explicitly that criticality is
marked by the vanishing of the{\em \ transverse} diffusion coefficient,
resulting in an estimate for $T_{c}(E)$ which increases with $E$, consistent
with MC data. On the quantitative side, we argue that fluctuations, largely
neglected in a series expansion such as ours, tend to increase the ratio $%
T_{c}(E)/T_{c}(0)$. In higher spatial dimensions, where fluctuations are
less relevant, agreement of series and MC data improves, borne out by our
results in $d=3$.

Quantitative comparisons aside, the approach presented here provides a
convenient analytic complement to MC simulations, since it gives direct
microscopic information about effective coarse-grained coupling constants
such as $\tau _{\bot }$ and $\tau _{\Vert }$ which appear in field theories.
It is computationally simpler than dynamic mean-field theory and easily
generalized \cite{long} to higher dimensions or other driven lattice models.
Thus, it can help to predict qualitative phase diagrams and formulate
effective field theories for a wide range of non-equilibrium steady states.

\section{Appendix}

Here, we give a few details of our calculations. First, we briefly review
the conventions of our Fourier transforms and then turn to the evaluation of
the integrals contributing to (\ref{I's_exp}).

Since our lattice consists of discrete points, we will let $(x,y)$ be all
pairs of integers. For our case, it is most convenient to define the
functions $U_{x,y}(k,p)$: 
\[
U_{x,y}(k,p)\equiv \frac{1}{2\pi }\exp i(kx+py) 
\]
with continuous $k,p\in $ $\left[ -\pi ,\pi \right] .$ The $U_{x,y}$ form a
complete orthonormal set: 
\[
\sum_{x,y}U_{x,y}(k,p)U_{x,y}(k^{\prime },p^{\prime })=\delta (k-k^{\prime
})\delta (p-p^{\prime }) 
\]
\[
\int dkdpU_{x,y}(k,p)U_{x^{\prime },y^{\prime }}(k,p)=\delta _{xx^{\prime }}%
\text{ }\delta _{yy^{\prime }}\text{.} 
\]
The Fourier transform is defined in the usual way,

\[
G(x,y)=\int S(k,p)\exp i(kx+py) 
\]
with inverse 
\[
S(k,p)=\sum_{x,y}G(x,y)\exp [-i(kx+py)] 
\]

Note that $S(k,p)$ is real, since $G(x,y)$ is even in both of its arguments.

Next, we turn to the integrals contributing to (\ref{I's_exp}). In order to
exhibit their properties succinctly, it will be helpful to write the
anisotropic Laplacian in the form 
\[
\Delta (k,p)\equiv A(1-\cos k)+B(1-\cos p) 
\]
where the values of interest, $A=2(1+\epsilon )$ and $B=4$ will be inserted
at the end. It is then easily seen from Eqns. (\ref{I's},\ref{LKP}), that
all integrals are of the general form 
\[
R_{ij}(A,B)\equiv \int \frac{(1-\cos k)^{i}(1-\cos p)^{j}}{\Delta (k,p)} 
\]
Clearly, $R_{ij}(A,B)=R_{ji}(B,A)$. Specifically, we need all pairs $(i,j)$
with $i,j=0,1,2,3$ except $(0,0)$. The calculations are simplified by a
series of identities, namely, 
\begin{eqnarray*}
&& \\
1 &=&\int \frac{\Delta (k,p)}{\Delta (k,p)}=AR_{10}+BR_{01} \\
1 &=&\int \frac{(1-\cos k)\Delta (k,p)}{\Delta (k,p)}=AR_{20}+BR_{11} \\
1 &=&\int \frac{(1-\cos p)\Delta (k,p)}{\Delta (k,p)}=AR_{11}+BR_{02} \\
\frac{3}{2} &=&\int \frac{(1-\cos k)^{2}\Delta (k,p)}{\Delta (k,p)}%
=AR_{30}+BR_{21} \\
1 &=&\int \frac{(1-\cos k)(1-\cos p)\Delta (k,p)}{\Delta (k,p)}%
=AR_{21}+BR_{12} \\
\frac{3}{2} &=&\int \frac{(1-\cos p)^{2}\Delta (k,p)}{\Delta (k,p)}%
=AR_{12}+BR_{03}\text{ .} \\
&&
\end{eqnarray*}
It is thus sufficient to compute $R_{i0}$ only. After performing the
elementary integral over $p$, the substitution $1-\cos k\equiv 2t$
generates, up to prefactors, the integral representation of Gauss'
hypergeometric function \cite{A+S}. Defining $z\equiv A/B$, we obtain 
\begin{eqnarray*}
R_{10} &=&\frac{2}{\pi B\sqrt{z}}F(\frac{1}{2},1;\frac{3}{2};-z) \\
R_{20} &=&\frac{8}{3\pi B\sqrt{z}}F(\frac{1}{2},2;\frac{5}{2};-z) \\
R_{30} &=&\frac{64}{15\pi B\sqrt{z}}F(\frac{1}{2},3;\frac{7}{2};-z) \\
&&
\end{eqnarray*}
We note in passing that all of these integrals can be expressed through
elementary functions, by reducing the hypergeometric functions down to $F(%
\frac{1}{2},1;\frac{3}{2},-z)=z^{-1/2}\arctan \sqrt{z}$. It is now
straightforward, if somewhat tedious, to compute $I_{1}$ and $I_{2}$, and
hence $\tau _{\Vert }$ and $\tau _{\bot }$, as functions of $z=(1+\epsilon
)/2$, in the region of interest $0\leq \epsilon \equiv \exp (-\beta E)$ $<1$%
. Both $I$'s are negative and decrease monotonically. Since the explicit
forms are not particularly illuminating, we quote a few representative
values in Table 1.\medskip 
\[
\]
\[
\begin{tabular}{|c|c|c|c|}
\hline
$\epsilon =$ & $~0$ & $~0.5$ & $~1$ \\ \hline
$~~~I_{1}/K=~$ & $~-0.844~$ & $~-0.931~$ & $-1$ \\ \hline
$~I_{2}/K=$ & $~-0.360~$ & $~-0.764~$ & $-1$ \\ \hline
\end{tabular}
\]

\begin{center}
Table 1.

Characteristic values for $I_{1}$ and $I_{2}$, in units of $K$, for
different field strengths.
\end{center}

\hspace{2.3in} 
\begin{eqnarray*}
&& \\
&& \\
&&
\end{eqnarray*}

\smallskip \medskip \noindent {\bf ACKNOWLEDGMENTS: }

BS wishes to thank M. Plischke, M. Grant and M. Zuckermann, as well as the
Physics Department of Simon Fraser University and the Centre for the Physics
of Materials at McGill University, for their warm hospitality while part of
this work was performed. We also thank Leah B. Shaw for valuable
discussions. Support from the US\ National Science Foundation through the
Division of Materials Research is gratefully acknowledged.

\end{document}